\documentclass[fleqn,english,aps,prd,preprintnumbers,showpacs,showkeys,nofootinbib,superscriptaddress,floatfix,tightenlines]{revtex4-2}
\usepackage{amsmath,amsfonts,amssymb,amscd,amsxtra,amsthm}
\usepackage{dsfont}
\usepackage{graphicx}  
\usepackage{epstopdf}
\usepackage{dcolumn}  
\usepackage{bm}          
\usepackage{slashed}
\usepackage[T1]{fontenc}
\usepackage[utf8]{inputenc}
\usepackage{booktabs}
\usepackage[normalem]{ulem} 
\usepackage[dvipsnames]{xcolor} 
\makeatletter
\usepackage{enumitem}
\usepackage{array}
\usepackage{tikz}
\usepackage{float}
\usepackage{multirow}
\usepackage{booktabs} 
\usepackage{slashed}

\renewcommand\sout{\bgroup \color{red} \ULdepth=-.5ex \ULset}

\makeatother
\usepackage{babel}
\begin{document}
\preprint{INHA-NTG-03/2024}
\title{\Large Two-pole structure of the $h_1(1415)$ axial-vector meson:
  resolving mass discrepancy}  

\author{Samson Clymton}
\email[E-mail: ]{sclymton@inha.edu}
\affiliation{Department of Physics, Inha University,
Incheon 22212, Republic of Korea }

\author{Hyun-Chul Kim}
\email[E-mail: ]{hchkim@inha.ac.kr}
\affiliation{Department of Physics, Inha University,
Incheon 22212, Republic of Korea }
\affiliation{School of Physics, Korea Institute for Advanced Study 
  (KIAS), Seoul 02455, Republic of Korea}
\date{\today}
\begin{abstract}
  We investigate isoscalar axial-vector mesons using a coupled-channel
  formalism. The kernel amplitudes are constructed from meson-exchange
  diagrams in the $t$- and $u$-channels, which are derived from
  effective Lagrangians based on hidden local symmetry. We incorporate
  six channels: $\pi\rho$, $\eta\omega$, $K\bar{K}^*$, $\eta\phi$,
  $\eta'\omega$, and $\eta'\phi$, and solve the off-shell coupled
  integral equations. We first discuss the dynamical generation of the
  $h_1(1170)$. The pole diagram for $h_1(1595)$ has a certain effect
  on the generation of $h_1(1170)$. We observe two poles at
  $(1387-i6)$ MeV and $(1452-i51)$ MeV, which exhibit a two-pole
  structure of the $h_1(1415)$ meson. This two-pole structure may
  resolve the discrepancy in the experimental data on the mass of
  $h_1(1415)$. The results show that the lower pole couples strongly
  to the $K\bar{K}^*$ channel, while the higher pole couples
  predominantly to the $\eta\phi$ channel. This provides insights into
  the nature of $h_1$ mesons and explains possible discrepancies in
  the mass of $h_1(1415)$. 
\end{abstract}
\maketitle

\section{Introduction}
Recently, the BESIII Collaboration reported new data on the excited
$h_1(1415)$ meson with quantum numbers $I^G(J^{PC})=0^-(1^{+-})$,
based on a partial-wave analysis of $J/\psi\to \gamma
\eta'\eta'$. They found its mass to be $m_{h_1(1415)}=(1384\pm
6_{-0}^{+9})$ MeV and width $\Gamma = (66\pm 10_{-10}^{+12})$
MeV~\cite{BESIII:2022zel}. These new results are consistent with the
earliest measurements of $h_1(1415)$\cite{Aston:1987ak}, which
reported $m_{h_1(1415)}=(1380\pm 20)$ MeV and $\Gamma=(80\pm 30)$ MeV 
from the $K^-p\to K_S^0 \bar{K}\pi\Lambda$ process. However, they
differ from several other experimental findings. The Crystal Barrel
experiment\cite{CrystalBarrel:1997kda} measured respectively its mass
and width as $(1440\pm 60)$ MeV and $(170\pm 80)$ MeV from
$p\bar{p}\to K_LK_S\pi^0\pi^0$. An earlier BESIII 
experiment~\cite{BESIII:2015vfb} found $m=(1412\pm 12)$ MeV and
$\Gamma=(84\pm52)$ MeV from $\chi_{1,2,J}\to\phi K\bar{K}\pi$. In
2018, BESIII~\cite{BESIII:2018ede} reported $m=(1423.2\pm 9.4)$ MeV
and $\Gamma=(90.3\pm27.3)$ MeV from $J/\psi\to \eta'K\bar{K}\pi$, with 
interference effects yielding $m=(1441.7\pm4.9)$ MeV and
$\Gamma=(111.5\pm12.8)$ MeV.  
Notably, the $K\bar{K}^*$ threshold energy
($E_{\mathrm{th}}^{K\bar{K}^*} \approx 1390$ MeV) lies between the 
$h_1(1415)$ masses reported in Refs.\cite{Aston:1987ak,
  BESIII:2022zel} and those in Refs.\cite{CrystalBarrel:1997kda,
  BESIII:2015vfb, BESIII:2018ede}.
It is crucial to understand the origin of these discrepancies. While
$h_1(1415)$ decays primarily into $K\bar{K^*}$, other channels near
its mass likely contribute as well. The lowest $h_1(1170)$ state
decays into $\pi\rho$, suggesting this channel may also affect
$h_1(1415)$ production. 

The Particle Data Group (PDG) classifies the $h_1$ mesons as
isoscalar $q\bar{q}$ states, i.e. $c_1(u\bar{u}+d\bar{d}) + c_2
s\bar{s}$ like $\eta$, $\eta'$, $\omega$, and $\phi$~\cite{PDG}. This
indicates that the $h_1$ mesons are considered to be orbitally excited 
states as the isoscalar pseudoscalar or vector mesons with the same
quark content. Similarly, the $a_1(1260)$ isovector axial-vector
meson is also regarded as the isovector $q\bar{q}$ state. However, a
series of studies suggests that the $a_1(1260)$ meson may be a
possible molecular state~\cite{Basdevant:1977ya, Roca:2005nm,
  Lutz:2003fm, Nagahiro:2011jn, Clymton:2022jmv}.  Very recently, we
have investigated the $b_1$ isovector axial-vector mesons,
demonstrating that they can be dynamically generated by considering
the four different channels, i.e. $\pi\omega$, $\eta\rho$, $\pi\phi$,
and $K\bar{K}^*$ channels. Interestingly, it was shown that the
$b_1(1235)$ arises from the $b_1(1306)$ and
$b_1(1356)$~\cite{Clymton:2023txd}, indicating that the
$b_1(1235)$ has a two pole structure. As will be shown in the current
work, the $h_1(1415)$ meson originates from the two poles: 
$h_1(1387)$ and $h_1(1452)$. In fact, various hadrons exhibit the
two-pole structures. For example, the $K_1(1270)$ may possibly be
regarded as the meson with the two-pole
structure~\cite{Roca:2005nm,Geng:2006yb}.  
Albaladejo et al.~\cite{Albaladejo:2016lbb} showed that the
$D^*(2400)$ arises as a two-pole structure, based on 
light pseudoscalar and $D$ meson interactions in the coupled-channel
formalism (see a recent review~\cite{Meissner:2020khl} for detailed
discussion). The two-pole structure is also found in the baryonic
sector: $\Lambda(1405)$ is now well established as the hyperon with
the two-pole structure~\cite{Oller:2000fj, Jido:2003cb,Xie:2023cej,PDG}.  

In the present work, we will show that the discrepancy in the
experimental data on the mass of $h_1(1415)$ is rooted in its two-pole
structure. To this end, we formulate the off-shell
coupled-channel formalism~\cite{Clymton:2022jmv, Clymton:2023txd,
  Kim:2023htt}, introducing six different channels, i.e. $\pi\rho$,
$\eta\omega$, $K\bar{K}^*$, $\eta\phi$, $\eta'\omega$, and $\eta'\phi$
of which the threshold energies lie below 2 GeV. 
We first construct the kernel amplitude corresponding to each channel,
using the meson-exchange diagrams. We compute the coupled
Blankenbecler-Sugar (BbS) equations, which are obtained from the
three-dimensional (3D) reduction of the  Bethe-Salpeter
equations~\cite{Blankenbecler:1965gx,   Aaron:1968aoz}. We have
already investigated how the $a_1(1260)$ and 
$b_1(1235)$ axial-vector mesons, $D_{s0}^*(2317)$ and $B_{s0}^*$
mesons, and the hidden charm pentaquark states are generated
dynamically within the same framework~\cite{Clymton:2022jmv,
  Clymton:2023txd, Kim:2023htt, Clymton:2024fbf}.
We will demonstrate that the $h_1(1170)$ and $h_1(1415)$ are
dynamically generated even without introducing the corresponding pole
diagrams. Remarkably, the two poles emerge in the second Riemann
sheet, which are related to the $h_1(1415)$ meson, of which one lies
just below the $K\bar{K}^*$ threshold, and the other is found to have
a larger mass than its threshold energy. 

The outline of the current work is sketched as follows: In
Section~\ref{sec:2}, we first explain how the kernel amplitude for
each channel can be constructed by using the Feynman diagrams based on
the effective Lagrangian. Then, we perform the partial-wave expansion
for the coupled BbS integral equations, so that we examine the
relevant partial waves with proper quantum numbers corresponding to
the $h_1$ mesons.  In Section~\ref{sec:3}, we discuss the results for
the $h_1$ mesons. We examine the role of each channel in producing
them dynamically. In particular, we focus on the two-pole structure of
the $h_1(1415)$ meson. The last section summarizes the present work. 

\section{General formalism\label{sec:2}}
\begin{figure}[htbp]
	\centering
	\includegraphics[scale=1.2]{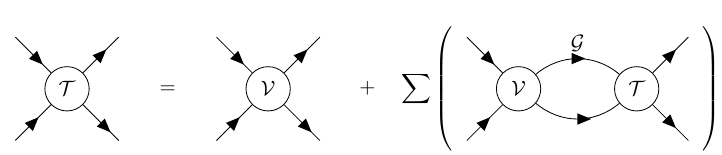}
	\caption{Graphical representation of the coupled integral
          scattering equation.}  
	\label{fig:1}
\end{figure}

The scattering amplitude is defined as 
\begin{align}
\mathcal{S}_{fi} = \delta_{fi} - i (2\pi)^4 \delta(P_f - P_i)
  \mathcal{T}_{fi}, 
\end{align}
where $P_i$ and $P_f$ stand for the total four momenta of the initial
and final states, respectively. The transition amplitude
$\mathcal{T}_{fi}$ for a two-body reaction can be derived from the
Bethe-Salpeter integral equations  
\begin{align}
\mathcal{T}_{fi} (p',p;s) =\, \mathcal{V}_{fi}(p',p;s) 
+ \frac{1}{(2\pi)^4}\sum_k \int d^4q 
\mathcal{V}_{fk}(p',q;s)\mathcal{G}_{k}(q;s) \mathcal{T}_{ki}(q,p;s),  
\label{eq:2}
\end{align}
where $p$ and $p'$ denote the relative four-momentum of the
initial and final states, respectively. $q$ is the momentum transfer
for the intermediate states in the center of mass (CM) 
frame. $s$ represents the square of the total energy, which is just
one of the Mandelstam variables, $s=P_i^2=P_f^2$. The coupled
integral equations given in Eq.~\eqref{eq:2} can be illustrated as in
Fig.~\ref{fig:1}. The summation $\Sigma$ represents the inclusion of
various coupled channels. 
To avoid the complexity due to the four-dimensional integral equations,
we make a 3D reduction. Among several methods for the 3D reduction, we
employ the BbS scheme~\cite{Blankenbecler:1965gx,
  Aaron:1968aoz}, which expresses the two-body propagator in the form of  
\begin{align}
	\mathcal{G}_k(q) =\;
  \delta\left(q_0-\frac{E_{k1}(\bm{q})-E_{k2}(\bm{q})}{2}\right)
	\frac{\pi}{E_{k1}(\bm{q})E_{k2}(\bm{q})}
  \frac{E_k(\bm{q})}{s-E_k^2(\bm{q})}.  
\label{eq:4}
\end{align}
Here, $E_k$ represents the total on-mass-shell energy of the
intermediate state, $E_k = E_{k1}+E_{k2}$, and $\bm{q}$
designates the three-momentum transfer of the intermediate
state. Utilizing Eq.~\eqref{eq:4}, we obtain the following coupled BbS
integral equations  
\begin{align}
	\mathcal{T}_{fi} (\bm{p}',\bm{p}) =\, \mathcal{V}_{fi}
	(\bm{p}',\bm{p})+\frac{1}{(2\pi)^3}\sum_k\int \frac{d^3
	\bm{q}}{2E_{k1}(\bm{q})E_{k2}(\bm{q})} \mathcal{V}_{fk}
	(\bm{p}',\bm{q})\frac{E_k(\bm{q})}{s-E_k^2(\bm{q})
	+i\varepsilon}\mathcal{T}_{ki}(\bm{q},\bm{p}),
	\label{eq:BS-3d}
\end{align}
where $\bm{p}$ and $\bm{p}'$ are the relative three-momenta of the 
initial and final states in the CM frame, respectively. In this
manner, the $T$ matrix can be generated, the entire Hilbert space
being considered with the off-shell components. 

Before we solve the coupled BbS integral equations, we need to
construct the kernel amplitudes $\mathcal{V}$. We compute
$\mathcal{V}_{fi}$ by using the effective Lagrangians for the
meson-meson interactions. Since the vector mesons are involved, we
consider hidden local symmetry, where the vector meson is considered
as a dynamic gauge boson~\cite{Fujiwara:1984mp, Bando:1987br}.  In
Ref.~\cite{Bando:1987br}, the effective interactions among vector,
vector, and pseudoscalar mesons were derived, based on the SU(3)
hidden local symmetry. The effective Lagrangians are then expressed as  
\begin{align}
  \mathcal{L}_{PPV} &= -ig_{PPV}\sqrt{2}\,\mathrm{Tr} 
  \left([P,\partial_\mu P]\,V^\mu\right),\cr 
\mathcal{L}_{VVV} &= ig_{VVV}\sqrt{2} \,
 \mathrm{Tr}\left((\partial_\mu V_\nu
- \partial_\nu V_\mu)V^\mu V^\nu\right),\cr 
\mathcal{L}_{PVV} &=-\frac{g_{PVV}}{m_V}\sqrt{2}\,\varepsilon^
{\mu\nu\alpha\beta}\mathrm{Tr}\left(\partial_\mu
V_\nu\partial_\alpha V_\beta P\right),
\label{eq:su3sym}
\end{align}
where $P$ and $V$ represent respectively the pseudoscalar and vector
matrices in flavor space  
\begin{align}
  P &= \begin{pmatrix}
    \frac{1}{\sqrt{2}} \pi^0+\frac{1}{\sqrt{6}}\eta_8+\frac{1}{
	\sqrt{2}}\eta_1 & \pi^+ & K^+\\
    \pi^- & -\frac{1}{\sqrt{2}} \pi^0+\frac{1}{\sqrt{6}}\eta_8
	+\frac{1}{\sqrt{2}}\eta_1 & K^0\\
    K^- & \bar{K}^0 & -\frac{2}{\sqrt{6}}\eta_8+\frac{1}{\sqrt{2}}
	\eta_1
  \end{pmatrix},\cr
  V &= \begin{pmatrix}
    \frac{1}{\sqrt{2}} \rho^0_\mu+\frac{1}{\sqrt{2}}\omega_\mu &
    \rho_\mu^+ & K_\mu^{*+}\\
    \rho_\mu^- & -\frac{1}{\sqrt{2}} \rho_\mu^0+\frac{1}{\sqrt{2}}
    \omega_\mu & K_\mu^{*0} \\
    K_\mu^{*-} & \bar{K}^{*0}_\mu & \phi_\mu
  \end{pmatrix}.
\end{align}
We assume the ideal mixing of the isoscalar vector meson singlet 
and octet. For $\eta$ and $\eta'$, we define them in terms of the 
pseudoscalar octet $\eta_8$ and singlet $\eta_1$: 
\begin{align}
	\eta = \eta_8 \cos\theta_P - \eta_1\sin\theta_P,\;\; \eta' = 
	\eta_8 \sin\theta_P + \eta_1\cos\theta_P,
\end{align}
with mixing angle $\theta_P = -17^\circ$ taken from 
Ref.~\cite{Amsler:1995td}. Note that the universal coupling constant is
given as $g=g_{PPV}=g_{VVV}$ due to the hidden local symmetry.

As done in the previous works~\cite{Clymton:2022jmv, Clymton:2023txd,
  Kim:2023htt}, we consider the five different isoscalar axial-vector
channels coupled to the $\pi\rho$ channel, which are relevant to the
$h_1$ mesons, i.e., the $\eta\omega$, $K\bar{K}^*$, $\eta\phi$,
$\eta'\omega$, and $\eta'\phi$ channels. Thus, the kernel matrix is
now expressed as  
\begin{align}
\mathcal{V} &= \begin{pmatrix}
	\mathcal{V}_{\pi\rho\to\pi\rho} & \mathcal{V}_{\eta\omega
	\to\pi\rho} & \mathcal{V}_{K\bar{K}^*\to\pi\rho} & 
	\mathcal{V}_{\eta\phi\to\pi\rho} & \mathcal{V}_{\eta'\omega
	\to\pi\rho} & \mathcal{V}_{\eta'\phi\to\pi\rho}\\ 
	\mathcal{V}_{\pi\rho\to\eta\omega} & \mathcal{V}_{\eta\omega
	\to\eta\omega} & \mathcal{V}_{K\bar{K}^*\to\eta\omega} & 
	\mathcal{V}_{\eta\phi\to\eta\omega} & \mathcal{V}_{\eta'\omega
	\to\eta\omega} & \mathcal{V}_{\eta'\phi\to\eta\omega}\\
	\mathcal{V}_{\pi\rho\to K\bar{K}^*} & \mathcal{V}_{\eta\omega
	\to K\bar{K}^*} & \mathcal{V}_{K\bar{K}^*\to K\bar{K}^*} & 
	\mathcal{V}_{\eta\phi\to K\bar{K}^*} & \mathcal{V}_{\eta'\omega
	\to K\bar{K}^*} & \mathcal{V}_{\eta'\phi\to K\bar{K}^*}\\
	\mathcal{V}_{\pi\rho\to \eta\phi} & \mathcal{V}_{\eta\omega\to
	\eta\phi} &	\mathcal{V}_{K\bar{K}^*\to \eta\phi} & 
	\mathcal{V}_{\eta\phi\to \eta\phi} & \mathcal{V}_{\eta'\omega
	\to \eta\phi} & \mathcal{V}_{\eta'\phi\to \eta\phi}\\
	\mathcal{V}_{\pi\rho\to \eta'\omega} & \mathcal{V}_{\eta\omega
	\to \eta'\omega} & \mathcal{V}_{K\bar{K}^*\to \eta'\omega} & 
	\mathcal{V}_{\eta\phi\to \eta'\omega} & \mathcal{V}_{\eta'\omega
	\to \eta'\omega} & \mathcal{V}_{\eta'\phi\to \eta'\omega}\\
	\mathcal{V}_{\pi\rho\to \eta'\phi} & \mathcal{V}_{\eta\omega
	\to \eta'\phi} & \mathcal{V}_{K\bar{K}^*\to \eta'\phi} & 
	\mathcal{V}_{\eta\phi\to \eta'\phi} & \mathcal{V}_{\eta'\omega
	\to \eta'\phi} & \mathcal{V}_{\eta'\phi\to \eta'\phi}\\
	\end{pmatrix}
\end{align}
\begin{figure}[htbp]
	\centering
	\includegraphics[scale=0.55]{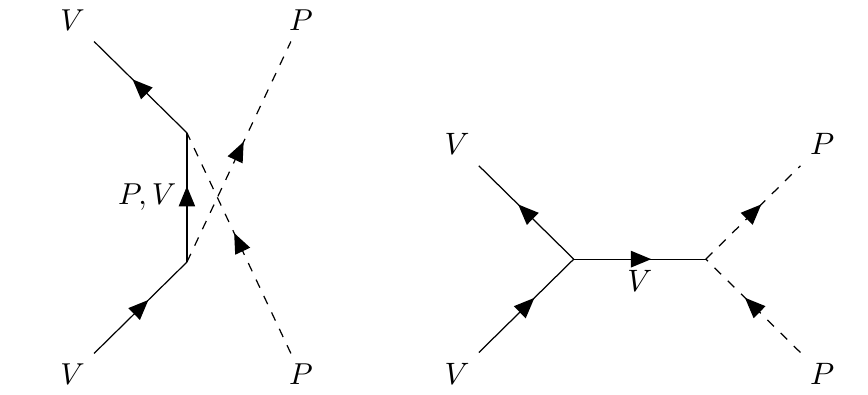}
	\caption{Meson-exchange diagrams in the $u$- and $t$-channels 
        are depicted in the left and right panels, respectively. $P$
        and $V$ stand for the pseudoscalar and vector mesons.}    
	\label{fig:2}
\end{figure}
Each matrix element $\mathcal{V}_{fi}$ is obtained from the
corresponding tree-level Feynman diagram illustrated in
Fig.~\ref{fig:2}. Note that we do not have any pole diagrams in the
$s$-channel, which indicates that the $h_1(1170)$ and $h_1(1415)$
mesons will be dynamically generated. However, the $h_1(1595)$ meson
requires the $s$-channel pole digram, which we will discuss later its
physical implications. The Feynman amplitudes for the tree-level
diagrams are evaluated from the effective Lagrangians  
given in Eq.\eqref{eq:su3sym}. Imposing flavor SU(3) symmetry for the
coupling constants, we find that the following kernel 
amplitudes become null:  
$\mathcal{V}_{\pi\rho\to\eta\phi}$, $\mathcal{V}_{\pi\rho\to\eta'
\phi}$, $\mathcal{V}_{\eta\omega\to\eta\phi}$, $\mathcal{V}_{\eta
\omega\to\eta'\phi}$, $\mathcal{V}_{\eta\phi\to\eta'\omega}$ and 
$\mathcal{V}_{\eta'\omega\to\eta'\phi}$. 

The amplitudes of the meson exchange diagrams shown in 
Fig.~\ref{fig:2} from left to right are written as
\begin{align}
	\mathcal{A}_{P}^{u}(\bm{p}',\bm{p}) =& -\mathrm{IS}\,g^2_{PPV}\,
	 F^2(\bm{p},\bm{p}')\,\left(2p_2-p_3\right)\cdot\epsilon^*
	\mathcal{P}(p_1-p_4)\,\left(2p_4-p_1\right)\cdot\epsilon,\\
	\mathcal{A}_{V}^{u}(\bm{p}',\bm{p}) =& -\mathrm{IS}\,\frac{g^2_
	{PVV}}{m_V^2}\,F^2(\bm{p},\bm{p}')\,\varepsilon_{\mu\nu\alpha\beta}
	\,p_3^\mu\epsilon^
	{*\nu}(p_3-p_2)^\alpha\,\mathcal{P}^{\beta\delta}(p_1-p_4)\,
	\varepsilon_{\gamma\sigma\eta\delta}p_1^\gamma 
	\epsilon^\sigma(p_1-p_4)^\eta,\\
	\mathcal{A}_{V}^{t}(\bm{p}',\bm{p}) =& -\mathrm{IS}\,g^2_{PPV}\,
	F^2(\bm{p},\bm{p}')\,\left(p_2+p_4\right)^\mu \mathcal{P}_{\mu\nu}
        (p_1-p_3)\cr
       &\times \left[(2p_1-p_3)\cdot\epsilon^*\epsilon^\nu+(2p_3-p_1)
	\cdot\epsilon\epsilon^{*\nu}-\epsilon\cdot\epsilon^*(p_1+p_3)^\nu
	\right],
\end{align}
where the IS factor is related to the corresponding SU(3) Clebsch-Gordan
coefficient and isospin factor. In Table~\ref{tab:1}, we list the
values of the IS factors for all relevant processes. For the coupling
constants, we use the values for the coupling constants:
$g_{PPV}^2/4\pi=0.72$  and $g_{PVV}^2/4\pi=1.88$ from the previous
works~\cite{Clymton:2022jmv,Clymton:2023txd}. The propagators for the
spin-0 and spin-1 mesons are expressed by  
  \begin{align}
	  \mathcal{P}(p) &= \frac{1}{p^2-m^2},\;\;\;
	  \mathcal{P}_{\mu\nu}(p) = \frac{1}{p^2-m^2}
	\left(-g_{\mu\nu}+\frac{p_\mu p_\nu}{m^2}\right), 
  \end{align}
where $m$ denotes the mass corresponding to the exchange meson. As
done in the previous works~\cite{Clymton:2022jmv,
  Clymton:2023txd}, we have turned off the energy-dependence in the
denominator of the propagator.
\begin{table}[htbp]
	\caption{\label{tab:1}The IS factors and 
	cutoff parameters ($\Lambda_0$) in unit of MeV for each
        reaction. }
  \begin{ruledtabular}
	\centering\begin{tabular}{lcccr}
	 Reaction & Exchange meson & 
	 Type & IS & $\Lambda_0$(MeV) 
	 \\\hline
	$\pi\rho \to \pi\rho$ & $\pi$ & $u$ & $-8$ & 600 \\
	                      & $\rho$ & $t$ & $-8$ & 600 \\
	                      & $\omega$ & $u$ & $4$ & 600 \\
	$\pi\rho \to \eta\omega$ & $\rho$ & $u$ & $-5.48$ & 1550 \\
	$\pi\rho \to K\bar{K}^*$ & $K$ & $u$ & $\sqrt{6}$ & 700 \\
	                         & $K^*$ & $t$ & $\sqrt{6}$ & 750 \\
	$\pi\rho \to \eta'\omega$ & $\rho$ & $u$ & $-4.24$ & 1600 \\
	$\eta\omega \to \eta\omega$ & $\omega$ & $u$ & $2.50$ & 600 \\
	$\eta\omega \to K\bar{K}^*$ & $K$ & $u$ & $-2.34$ & 600 \\
	                            & $K^*$ & $t$ & $-2.34$ & 700 \\
	$\eta\omega \to \eta'\omega$ & $\omega$ & $u$ & $1.94$ & 600 \\
	$K\bar{K}^* \to K\bar{K}^*$ & $\rho$ & $t$ & $-3$ & 600 \\
	                            & $\omega$ & $t$ & $-1$ & 600 \\
	                            & $\phi$ & $t$ & $-2$ & 1400 \\
	$K\bar{K}^* \to \bar{K}K^*$ & $\pi$ & $u$ & $3$ & 600 \\
	                            & $\eta$ & $u$ & $2.74$ & 600 \\
	                            & $\rho$ & $u$ & $-3$ & 600 \\
	                            & $\omega$ & $u$ & $-1$ & 600 \\
	                            & $\phi$ & $u$ & $-2$ & 1400 \\
	$K\bar{K}^*\to\eta\phi$ & $K$ & $u$ & $-3.31$ & 600 \\
	                        & $K^*$ & $t$ & $-3.31$ & 1710 \\
	$K\bar{K}^*\to\eta'\omega$ & $K$ & $u$ & $0.72$ & 600 \\
	                        & $K^*$ & $t$ & $0.72$ & 700 \\
	$K\bar{K}^*\to\eta'\phi$ & $K$ & $u$ & $1.01$ & 600 \\
	                        & $K^*$ & $t$ & $1.01$ & 1700 \\
	$\eta\phi \to \eta\phi$ & $\phi$ & $u$ & $3.00$ & 600 \\
	$\eta\phi \to \eta'\phi$ & $\phi$ & $u$ & $-3.87$ & 600 \\
	$\eta'\omega \to \eta'\omega$ & $\omega$ & $u$ & $1.50$ & 600 \\
	$\eta'\phi \to \eta'\phi$ & $\phi$ & $u$ & $5.00$ & 600 \\
    \end{tabular}
  \end{ruledtabular}
\end{table}

Since hadrons have finite sizes, we introduce a form factor
at each vertex. We employ the following
parametrization~\cite{Kim:1994ce}: 
\begin{align}
	F(\bm{p},\bm{p}') =
  \left(\frac{n\Lambda^2-m^2}{n\Lambda^2+\bm{p}^2+\bm{p}'^2}\right)^n,  
\label{eq:13} 
  \end{align}
where $n$ is the power of the form factor. The form given in
Eq.~\eqref{eq:13} has a notable advantage: the value of $\Lambda$
remains constant regardless of change in $n$. As $n$ approaches
infinity, Eq.~\eqref{eq:13} converges to a Gaussian 
form. In most cases, we use $n=1$. However, we need to use $n=2$ for
vector-meson exchange, because they have stronger momentum
dependence. While the cut-off masses $\Lambda$ in Eq.~\eqref{eq:13}
are experimentally unknown for the current hadronic processes, we
minimize associated uncertainties using the following strategy as done
in the previous works~\cite{Clymton:2022jmv, Clymton:2023txd,
  Kim:2023htt}: we choose $\Lambda$ by adding approximately
$(500-700)$ MeV to the corresponding masses of the exchange
meson. Consequently, we define the cutoff mass as
$\Lambda=\Lambda_0+m$. We set $\Lambda_0$ to 
be $600$ MeV for most cases, as listed in Table.~\ref{tab:1}. However,
to fit the total cross section for $\pi\rho$ scattering, we need to
use larger values of $\Lambda_0$ for some of the vector-meson exchange 
diagrams, as shown in Table~\ref{tab:1}.

To compute the coupled integral equations, we utilize a partial 
wave decomposition, transforming the equation into a one-dimensional 
integral equation as follows
\begin{align}
	\mathcal{T}^{J(fi)}_{\lambda'\lambda} (\mathrm{p}',\mathrm{p}) = 
	\mathcal{V}^{J(fi)}_{
	\lambda'\lambda} (\mathrm{p}',\mathrm{p}) + \frac{1}{(2\pi)^3}
	\sum_{k,\lambda_k}\int
	\frac{\mathrm{q}^2d\mathrm{q}}{2E_{k1}E_{k2}}
	\mathcal{V}^{J(fk)}_{\lambda'\lambda_k}(\mathrm{p}',
	\mathrm{q})\frac{E_k}{
	s-E_k^2} \mathcal{T}^{J(ki)}_{\lambda_k\lambda}
	(\mathrm{q},\mathrm{p}).
	\label{eq:BS-1d}
\end{align}
Here, $\lambda'$, $\lambda$ and $\lambda_k$ denote the relative 
helicity of the final, initial and intermediate two-body states, 
respectively. Their corresponding momenta are represented by 
$\mathrm{p}'$, $\mathrm{p}$ and $\mathrm{q}$, respectively. The
partial-wave component is given by
\begin{align}
	\mathcal{V}^{J(fi)}_{\lambda'\lambda}(\mathrm{p}',\mathrm{p}) = 
	2\pi \int d(
	\cos\theta) \,d^{J}_{\lambda',\lambda}(\theta)\,\mathcal{V}^{fi}_{
	\lambda'\lambda}(\mathrm{p}',\mathrm{p},\theta),
\end{align}
where $\theta$ stands for the scattering angle, and $d^J_{\lambda',
\lambda}$ are the Wigner $d$-matrices.

To obtain the transition amplitudes numerically, it is crucial to 
deal with the singularities in the BbS propagator. The
one-dimensional integral, free from energy singularities, takes the
form  
\begin{align}
	\mathcal{T}^{fi}_{\lambda'\lambda} (\mathrm{p}',\mathrm{p}) = 
	\mathcal{V}^{fi}_{
	\lambda'\lambda} (\mathrm{p}',\mathrm{p}) + \frac{1}{(2\pi)^3}
	\sum_{k,\lambda_k}\left[\int_0^{\infty}d\mathrm{q}
	\frac{\mathrm{q}E_k}{E_{k1}E_{k2}}\frac{\mathcal{F}(\mathrm{q})
	-\mathcal{F}(\tilde{\mathrm{q}}_k)}{s-E_k^2}+ \frac{1}{2\sqrt{s}}
	\left(\ln\left|\frac{\sqrt{s}-E_k^{\mathrm{thr}}}{\sqrt{s}
	+E_k^{\mathrm{thr}}}\right|-i\pi\right)\mathcal{F}
	(\tilde{\mathrm{q}}_k)\right],
	\label{eq:BS-1d-reg}
\end{align}
where $\tilde{\mathrm{q}}_k$ represents the momentum when 
$E_{k1}+E_{k2}=\sqrt{s}$ and $\mathcal{F}(\mathrm{q})$ is defined as 
\begin{align}
	\mathcal{F}(\mathrm{q})=\frac{1}{2}\mathrm{q}\,
	\mathcal{V}^{fk}_{\lambda'\lambda_k}(\mathrm{p}',
	\mathrm{q})\mathcal{T}^{ki}_{\lambda_k\lambda}(\mathrm{q},
	\mathrm{p}).
\end{align} 

Regularization is applied only when the total energy $\sqrt{s}$ 
surpasses the threshold energy of the $k$-th channel, denoted as 
$E_k^{\mathrm{thr}}$. Notably, the transition amplitudes derived from
these equations can be analytically continued to the complex
energy plane directly, as the energy singularities have been
eliminated. Moreover, Eq.~\ref{eq:BS-1d-reg} enables the use  
of the matrix inversion method to compute the transition
amplitudes. The partial-wave transition amplitudes are then obtained
by transforming the helicity basis into the LSJ basis. For the
specific case of pseudoscalar and vector meson scattering, this
transformation is represented as 
\begin{align}
	\mathcal{T}^{JS}_{L'L} = \frac{\sqrt{(2L+1)(2L'+1)}}{2J+1}
	\sum_{\lambda'\lambda} \left(L'0S\lambda'|J\lambda'\right)
	\left(1\lambda' 00|1\lambda'\right)\left(L0S\lambda|J\lambda\right)
	\left(1\lambda 00|S\lambda\right)\mathcal{T}^{J}_{\lambda',
	\lambda}.
\end{align}
Here, $(j_1m_1j_2m_2|JM)$ denotes the Clebsch-Gordan coefficient.

\begin{figure}[htbp]
	\centering
	\includegraphics[scale=0.25]{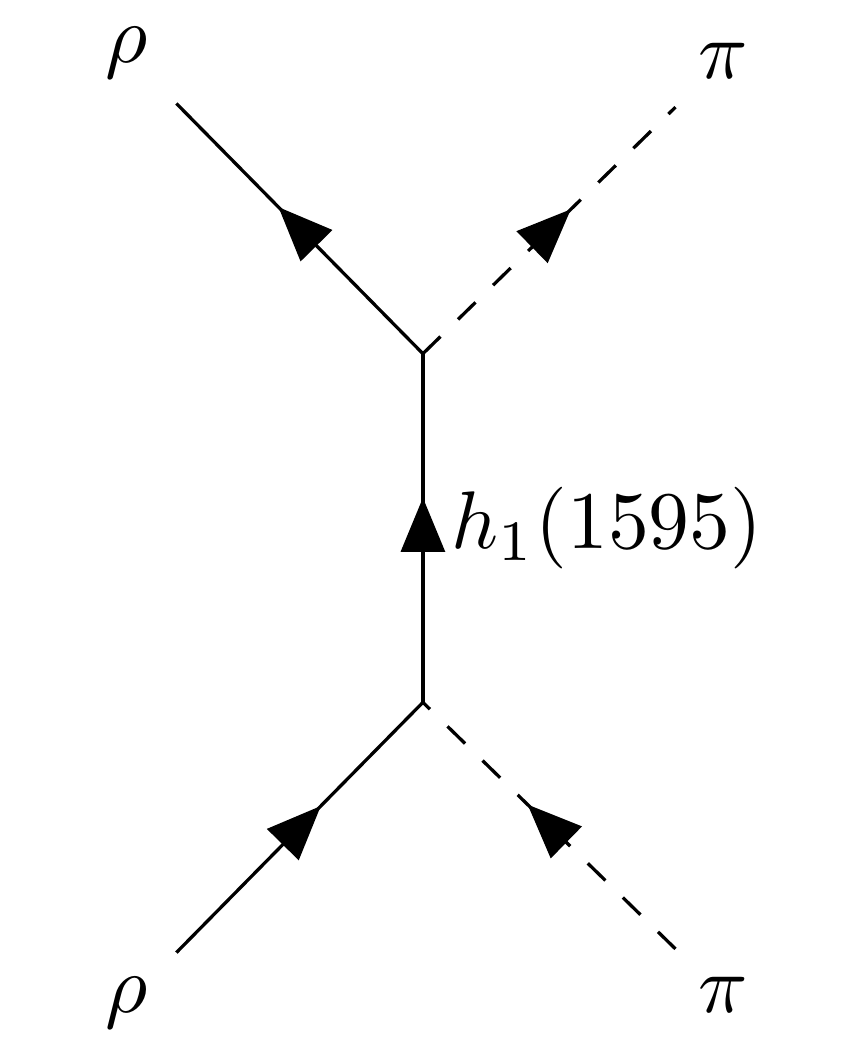}
	\caption{$h_1$(1595) pole diagram in the $\pi\rho$ elastic 
	scattering.}  
	\label{fig:3}
\end{figure}
So far, we have focused on using only the $t$ and $u$ channel  
exchange diagrams. However, it is necessary to include the pole
diagram for the $h_1(1595)$ meson in the $s$ channel. It is   
essential to improve the results in comparison with the experimental
data on $\pi\rho$ scattering. A similar approach was demonstrated in
the case of $\pi\pi$ scattering~\cite{Lohse:1990ew}, where the
explicit inclusion of the $f_0$(1400) pole was necessary to reproduce
the experimental phase shift of the scalar-isoscalar channel in  
the energy region below 2 GeV. The presence of $f_0$(1400) influenced
both the high-energy region and the structure below 1 GeV. Similarly, we
incorporate the $s$-channel pole diagram for $h_1(1595)$ in $\pi\rho$
scattering, as illustrated in Fig.~\ref{fig:3}. The interaction
vertices are determined by the following effective Lagrangian  
  \begin{align}
	\mathcal{L}_{h_1\pi\rho} = \frac{g_{h_1\pi\rho}}{m_{h_1}^0}
	\left(\partial_\mu\rho_\nu-\partial_\nu\rho_\mu\right)
	\pi\partial^\mu h_1^\nu.
  \end{align}
The bare mass and coupling of $h_1$(1595) are set to be  
$m_{h_1(1595)}^{(0)}=1320$ MeV and
$g_{h_1(1595)\pi\rho}^2/4\pi=0.54$. Additionally,  
we employ the form factor in the $s$ channel 
\begin{align}
	F_s(\bm{p}) = \left(\frac{\Lambda^2+(m_{h_1}^0)^2}{\Lambda^2
	+\bm{p}^2}\right)^4,
\end{align}
to ensure that the contribution of the pole diagram in the high 
momentum region is negligible. Consequently, the pole diagram has a 
small impact on the dynamical generation of the $h_1$ meson. We set 
the cutoff mass, denoted as $\Lambda$, to be $1920$ MeV, following the 
previously mentioned rule. As a result, the dressed mass and width of 
$h_1$(1595) evaluated in the complex energy plane match the PDG 
average value~\cite{PDG}. We found the $u$-channel  
contribution to the transition amplitude to be negligible and omitted it  
for simplicity. The $t$-channel diagram is not allowed due to $G$
parity and isospin symmetry. The necessity of including the pole
diagram suggests that the $h_1(1595)$ may be predominantly a 
genuine $q\bar{q}$ state.  

\section{Results and discussions \label{sec:3}}
\subsection{$h_1$ resonances}
The ground state of the $h_1$ axial-vector meson was first observed in 
the $3\pi$ mass spectra of the charge exchange reaction $\pi^- p \to 
\pi^+\pi^-\pi^0 n$~\cite{Dankowych:1981ks}. Its existence was later 
confirmed by only one other experiment using the same
reaction~\cite{Ando:1990ti}. According to the PDG, there
are two excited states of $h_1$ meson: $h_1$(1415) and
$h_1$(1595)~\cite{PDG}. The latter was observed in the $\eta\omega$
mass distribution~\cite{BNL-E852:2000poa}, while the former  
exhibits an intriguing structure similar to the renowned $\Lambda$
(1405). Initially, referred to as $h_1$(1380), 
subsequent experimental studies determined its mass to be
approximately $1.42-1.44$ GeV, leading to its renaming as
$h_1$(1415). However, a very recent experiment again suggests that the
mass of $h_1$(1415) is roughly $1.38$ GeV. Currently, there is no
explanation for the conflicting mass measurements of $h_1$(1415).
Interestingly, $h_1$(1380) is located just below the $K\bar{K}^*$
threshold, whereas the renamed $h_1$(1415) is found to be above
 $K\bar{K}^*$ threshold. This implies that $h_1(1380)$ ($h_1(1415)$)
 may have a two-pole structure.  In this work, we will examine each
 $h_1$ meson within the coupled-channel framework. Specifically, we
 will demonstrate that the conflicting mass measurements of
 $h_1$(1415) can be explained by a two-pole structure. 

In Ref.~\cite{Dankowych:1981ks}, two axial-vector resonances, $a_1$ 
and $h_1$, were identified in the charge exchange
reaction. Previously, we analyzed experimental data on the former one,
and in this current study, we extend our analysis to the  
latter one, keeping values of the parameters the same as the earlier 
investigation~\cite{Clymton:2022jmv}. The assumption of mixing 
between $\eta$ and $\eta'$ alters the isospin value of $\eta$ exchange 
in the $K\bar{K}^*$ elastic channel. However, the mixing has 
a negligible effect on the previous results for the $a_1$ meson, given
the fact that $\eta$ exchange barely contributes to $K\bar{K}^*$ elastic 
scattering.

\begin{figure}[htbp]
	\centering
	\includegraphics[scale=0.8]{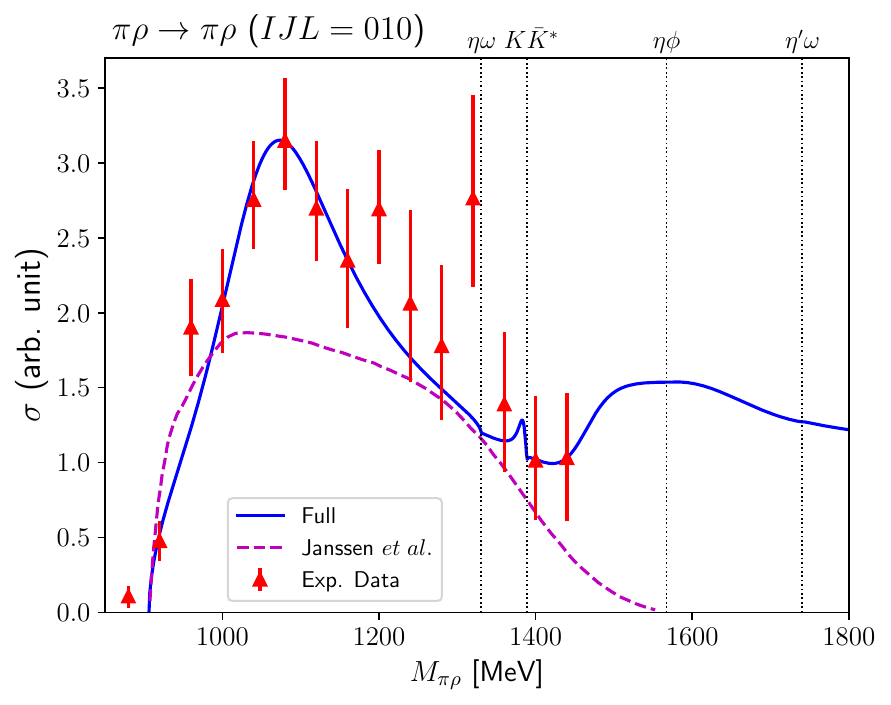}
	\caption{Total cross section for $\pi\rho$ scattering in the
          $IJL=010$ channel as a function of the $\pi\rho$ invariant 
	  mass. The solid curve depicts the present results, whereas
          the dashed one draws that from
          Ref.~\cite{Janssen:1996kx}. The experimental data are taken
          from Ref.~\cite{Dankowych:1981ks}.}    
	\label{fig:4}
\end{figure}
To describe the experimental data in the isoscalar channel, we relate
the total cross section to the transition amplitude $\mathcal{T}$ using a 
constant factor $C$, expressed as
\begin{align}
\sigma \equiv -C\, \mathrm{Im}\left[\mathcal{T}_{\pi\rho}(M_{\pi\rho})
\right],
\end{align}
where $C$ has a different value from that in the isovector channel
due to the absorption of the isospin factor into $C$. The pole diagram
for $h_1(1595)$ comes into essential play in enhancing the width of
$h_1(1170)$, resulting in good agreement with the experimental
data. The total cross section for $\pi\rho$ scattering in the
isoscalar channel clearly reveals the resonance of the $h_1$(1170)
meson, of which the mass and width are determined to be $m_{h_1(1170)}
= (1.19\pm0.06)$ GeV and $\Gamma = (0.32\pm 0.05)$ GeV,
respectively~\cite{Dankowych:1981ks}. 
The present results describe the experimental data well. We also
compare them with those in Ref.~\cite{Janssen:1996kx} that utilized a
single $\pi\rho$ channel including explicitly the $h_1$(1170)
pole diagram. As shown in Fig.~\ref{fig:4}, the single channel with
the explicit pole of the $h_1$(1170) is insufficient to describe the 
experimental data, specifically the $h_1$(1170) peak structure. In
contrast, we find that the $h_1(1595)$ pole diagram has an important
contribution to explain the $h_1$(1170) resonance. Moreover, 
various coupled channels play essential roles in dynamically
generating the $h_1$ (1170) resonance. The dynamical generation of
$h_1$ (1170) has also been observed in Ref.~\cite{Roca:2005nm}.  
So, we conclude that the $h_1$(1170) does not solely originate from the
$q\bar{q}$ state but contains a substantial component of the molecular
state. Furthermore, we predict a small peak structure near the
$K\bar{K}^*$ threshold, which is barely seen in the experimental
data.

\begin{figure}[htbp]
	\centering
	\includegraphics[scale=0.9]{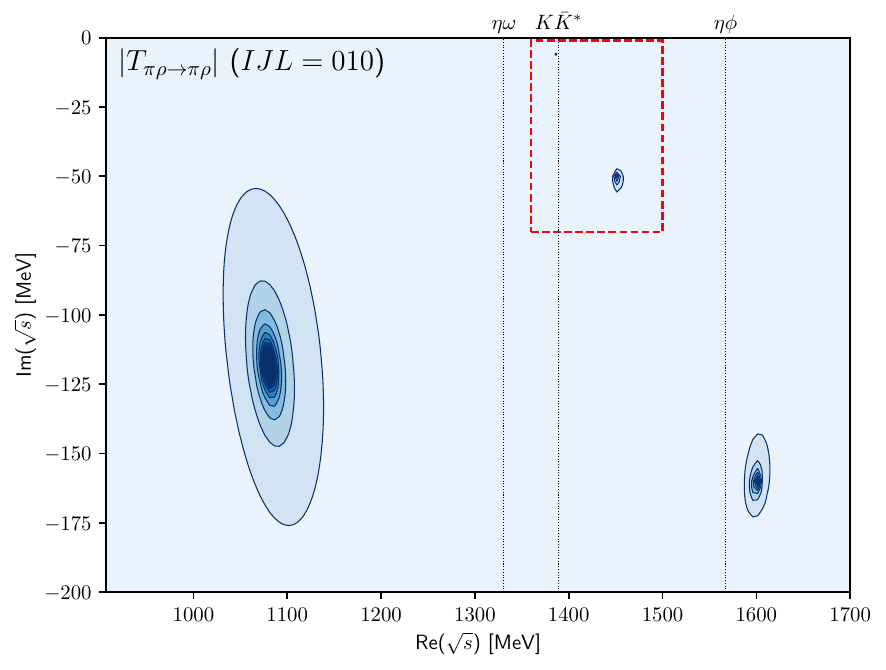}
	\caption{Contour plot of the modulus of the $\pi\rho$
	transition amplitude in the complex $\sqrt{s}$ plane. Four
        poles are found in the second Riemann sheet. The contour plot
        in the red rectangle is enlarged and shown in Fig.~\ref{fig:6}.}
	\label{fig:5}
\end{figure}
To identify dynamically generated resonances in the current approach,
we examine the $T$ amplitude in the second Riemann sheet.   
Since the $T$ amplitude generated by Eq.~\eqref{eq:BS-1d-reg} is
a meromorphic function in the complex energy plane, we can directly
identify the $h_1$ resonances in the second Riemann sheet. In
Fig.~\ref{fig:5}, the contour plot of the modulus of $T$ clearly
exhibits the existence of four poles. The first pole below the
$\eta\omega$ threshold is positioned at $(1080-i118)$ MeV, so its mass
and width are respectively given as $1080$ MeV and $236$ MeV.
The value of the width given in the PDG is $(375\pm 35)$
MeV~\cite{PDG}, which was taken from the fit by using the Bowler
model. Compared with it, the current result is smaller than the
empirical data. In Ref.~\cite{Roca:2005nm}, the corresponding pole for
$h_1$(1170) was found at $\sqrt{s}=(919-i17)$ MeV, which deviates
significantly from the experimental data.
The second and third poles are located at $(1387-i6)$ MeV and
$(1452-i51)$ MeV, respectively. These two poles are related to 
$h_1(1415)$, once called as $h_1(1380)$. Since the second pole lies
just below the $K\bar{K}^*$ threshold, it may be considered as the
$K\bar{K}^*$ molecular state in the isoscalar channel. Interestingly,
its width is very small. The third pole is located at about 60 MeV above 
the $K\bar{K}^*$ threshold, and its width is 102 MeV. While the
average value from the PDG is given as $(78\pm 11)$ MeV, the
experimental data on its width ranges from $(66\pm 10_{-10}^{+12})$
MeV~\cite{BESIII:2022zel}  to $(170\pm 80)$
MeV~\cite{CrystalBarrel:1997kda}. Thus, the width of $h_1(1415)$
should be measured more precisely. If $h_1(1415)$ has a two-pole
structure, its width may be determined more precisely. 

Finally, the fourth pole is observed at $(1603-i158)$ MeV that
corresponds to the $h_1(1595)$ meson, placed above the $\eta\phi$
threshold. Note that we have included the pole diagram for $h_1(1595)$
in the $s$ channel with the bare mass $m_{h_1(1595)}^{(0)}=1320$
MeV. It is dressed to be 1603 MeV,   
as the pole diagram for $h_1(1595)$ is ``\emph{renormalized}'' by
the coupled integral equations. When considering only the
$\pi\rho$ elastic channel, the dressed mass is smaller than the bare
one. However, the pole diagram gains additional mass after being
coupled to all channels. A similar effect is observed in $\pi\pi$
elastic scattering, where the scalar-isoscalar meson $f_0$(1400) has a
bare mass of $1300$ MeV with the scalar coupling, and after coupling to
other channels, the physical mass becomes $1400$
MeV~\cite{Lohse:1990ew}. While a recent study shows that the
$h_1$(1595) is a ground state of a light tetraquark
state~\cite{Zhao:2021jss}, the current result suggests that the
structure of the $h_1(1595)$ has a large component of the $q\bar{q}$,
being dressed by various channels. 

\begin{figure}[htbp]
	\centering
	\includegraphics[scale=0.8]{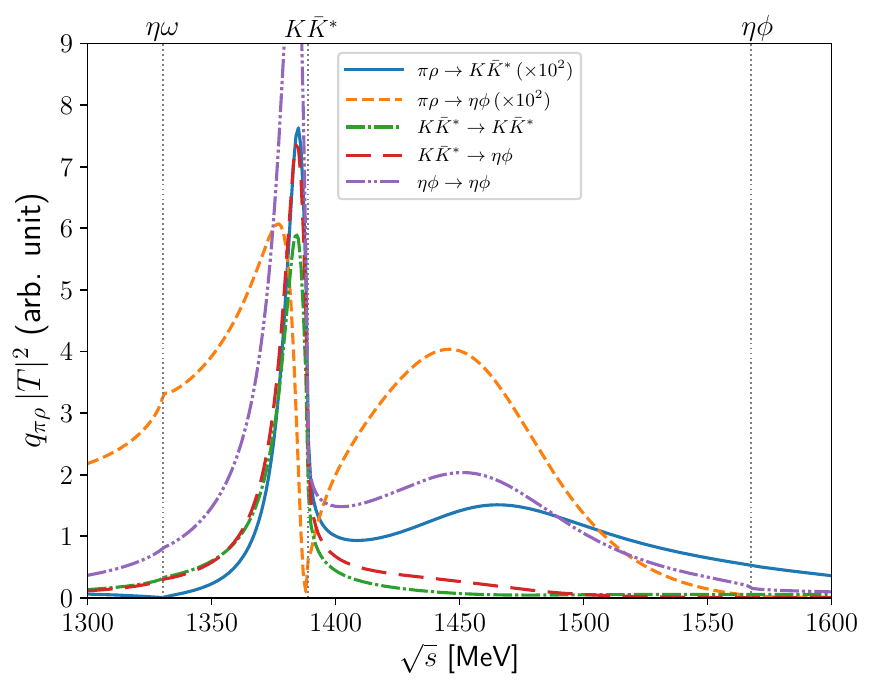}
	\caption{Results for $q_{\pi\rho}\left|T\right|^2$ for various
          channels. $q_{\pi\rho}$ is the magnitude of the momentum of 
          the $\pi\rho$ system.} 
	\label{fig:6}
\end{figure}

The coupling strengths of the $h_1$ resonances can be derived from the
residues of the $\mathcal{T}$ matrix, defined as 
\begin{align}
	\mathcal{R}_{a,b} &:= \lim_{s\to s_R} \left(s-s_R\right)
	\mathcal{T}_{a,b}/4\pi=g_ag_b.
\end{align}
It is impossible to determine the signs of the coupling
strengths, so we choose the coupling strengths to the $\pi\rho$  
channel to be positive. Then, we can determine the relative signs for
other coupling strengths. Table~\ref{tab:2} lists the results for the
coupling strengths to all possible channels.
\begin{table}[htp]
	\caption{\label{tab:2}Coupling strengths in units of GeV of the
		  $h_1$ resonance to the $S$ and $D$ wave states in
                  the six different channels. Note that we fix the
                  relative signs of the coupling strengths choosing
                  the coupling strengths to the $\pi\rho$ channel to be
                  positive.}  
	\begin{ruledtabular}
	\begin{tabular}{lcccr}
	$\sqrt{s_R}$ [MeV] & $1080-i118$ & $1387-i6$ & $1452-i51$ & 
	$1603-i158$\\
	\hline
	$g_{\pi\rho}$($S$-wave)    & $ 4.23-i1.80$ & $ 0.43+i0.20$ & 
	$ 0.85-i1.37$ & $ 2.81+i0.28$\\ 
	$g_{\pi\rho}$($D$-wave)    & $ 0.48-i0.42$ & $ 0.18+i0.17$ & 
	$ 1.09+i0.11$ & $ 1.51-i0.15$\\ 
	$g_{\eta\omega}$($S$-wave) & $-1.05+i0.87$ & $ 1.74+i0.49$ & 
	$ 1.69+i1.49$ & $ 2.42-i2.01$\\
	$g_{\eta\omega}$($D$-wave) & $-0.12+i0.03$ & $-0.09-i0.02$ & 
	$-0.19-i0.01$ & $-0.77+i0.80$\\
	$g_{K\bar{K}^*}$($S$-wave) & $-0.47+i0.12$ & $-4.23-i0.95$ & 
	$-1.20-i2.74$ & $ 1.92-i0.80$\\
	$g_{K\bar{K}^*}$($D$-wave) & $-0.04-i0.01$ & $ 0.00+i0.01$ & 
	$-0.21+i0.50$ & $-1.78+i0.32$\\
	$g_{\eta\phi}$($S$-wave)   & $-3.96+i0.79$ & $ 4.84+i1.23$ & 
	$ 8.83+i1.52$ & $ 0.04+i4.12$\\ 
	$g_{\eta\phi}$($D$-wave)   & $-0.51-i0.03$ & $-0.21-i0.06$ & 
	$-0.22-i0.15$ & $-0.42+i0.43$\\ 
	$g_{\eta'\omega}$($S$-wave)& $-0.74+i0.69$ & $-0.06+i0.19$ & 
	$ 1.26+i0.23$ & $ 2.51-i0.95$\\ 
	$g_{\eta'\omega}$($D$-wave)& $-0.20+i0.12$ & $ 0.01-i0.02$ & 
	$-0.11-i0.04$ & $-0.15-i0.09$\\ 
	$g_{\eta'\phi}$($S$-wave)  & $ 0.66-i0.13$ & $-1.52-i0.31$ & 
	$-1.83-i0.67$ & $ 0.30-i0.67$\\ 
	$g_{\eta'\phi}$($D$-wave)  & $ 0.16-i0.01$ & $ 0.23+i0.05$ & 
	$ 0.23+i0.11$ & $-0.05+i0.05$\\ 
	\end{tabular}
	\end{ruledtabular}
      \end{table}
The first resonance, $h_1(1170)$, couples most strongly to the
$\pi\rho$ channel, which is in line with the experimental 
observations of the $h_1(1170)\to \pi\rho$
decay~\cite{Dankowych:1981ks, OmegaPhoton:1983vde,  Ando:1990ti}.  
The next strongest coupling comes from the $\eta \phi$
channel. Considering that the $\eta \phi$ threshold  energy is
around 1570 MeV, it is remarkable that the $\eta \phi$ channel is the
second most dominant one for $h_1(1170)$. This indicates that the
$h_1(1170)$ contains a significant $s\bar{s}$ component, resembling
the case of dynamically generated $a_1$(1260) and
$b_1$(1235)~\cite{Clymton:2022jmv, Clymton:2023txd}. The $\eta\phi$
channel couples most strongly to the second and third
resonances, located at $(1387-i6)$ MeV and $(1452-i51)$ MeV, 
which are considered to be $h_1(1415)$. As depicted in
Fig.~\ref{fig:6}, one can clearly observe the strongest peak structure 
below the $K\bar{K}^*$ threshold in the squared modulus of the
$\eta\phi\to \eta\phi$ transition amplitude multiplied by
$q_{\pi\rho}$, $q_{\pi\rho}|T|^2$, where $q_{\pi\rho}$ is the
magnitude of the momentum of the $\pi\rho$ system. 
Interestingly, we also have nonvanishing $\pi\rho\to\eta\phi$
transition amplitudes, as shown in Fig.~\ref{fig:6}. 
Although we impose the flavor SU(3) symmetry, which results in the
absence of the $\pi\rho\to \eta\phi$ kernel amplitude, the
$\pi\rho\to\eta\phi$ transition amplitude is generated through the
$K\bar{K}^*$ intermediate states, revealing a resonance
structure below the $K\bar{K}^*$ threshold.  
As expected, the second resonance has a strong coupling strength to
the $K\bar{K}^*$ channel (see also the dash-dotted curve in
Fig.~\ref{fig:6}). On the other hand, the coupling strengths to 
the $S$-wave $\eta\omega$, $K\bar{K}^*$, $\eta'\omega$, and
$\eta'\phi$ channels are of the same order. It is worth noting that
the $D$-wave $\pi\rho$ channel has a sizable coupling strength. 
As for the $h_1(1595)$, many channels are strongly coupled to it. As
explained eralier, the mass of $h_1(1595)$ would have been less than
the bare mass had we considered the $\pi\rho$ channel only. As shown
in Table~\ref{tab:2}, the $\pi\rho$, $\eta\omega$, $K\bar{K}^*$, and
$\eta'\omega$ contribute to the generation of the $h_1(1595)$ meson.  

\subsection{Two-pole structure of the $h_1(1415)$}
\begin{table}[htbp]
	\caption{\label{tab:3}The experimental results on the mass and 
	width of $h_1$(1415) from various collaborations.} 
	\begin{ruledtabular}
	\begin{tabular}{lcccr}
	Collaboration & Reference & Process & Mass [MeV] & Width [MeV]\\
	\hline
	LASS & \cite{Aston:1987ak} & $K^{-}p\to K_S^0\bar{K}\pi\Lambda$ & 
	$1380\pm20$ & $80\pm 30$\\
	Crystal Barrel & \cite{CrystalBarrel:1997kda} & $p\bar{p}\to K_L 
	K_S\pi^0\pi^0$ & $1440\pm60$ & $170\pm 80$\\
	BESIII & \cite{BESIII:2015vfb} & $\chi_{1,2,J}\to\phi K\bar{K}
	\pi$ & $1412\pm 12$ & $84\pm 52$\\
	BESIII & \cite{BESIII:2018ede} & $J/\psi\to\eta' K\bar{K}\pi$ & 
	 $1423.2\pm 9.4$ & $90.3\pm 27.3$\\
	 &  &  & $1441.7\pm 4.9^{(*)}$ & $111.5\pm 12.8^{(*)}$\\
	BESIII & \cite{BESIII:2022zel} & $J/\psi\to\gamma\eta' \eta'$ & 
	 $1384\pm6^{+9}_{-0}$ & $66\pm 10^{+12}_{-10}$\\
	\end{tabular}
	\end{ruledtabular}
\raggedright
{\footnotesize ${}^{(*)}${Interference effect is considered}
}
\end{table}
Table~\ref{tab:3} presents the mass of the $h_1(1415)$ as measured by
various experiments. The LASS Collaboration initially detected a
signal of $h_1(1415)$ in the $K\bar{K}\pi$ system using the $K^{-}p\to
K_S^0\bar{K}\pi\Lambda$ reaction~\cite{Aston:1987ak}. The data
exhibited an enhancement near 1.4 GeV, diminishing rapidly as the
energy increased. Using the Breit-Wigner parameterization, the mass and
width were determined to be $M=1380\pm 20$ MeV and $\Gamma= 80\pm 30$
MeV, respectively. 
Subsequent experiments, however, reported different results. The
Crystal Barrel Collaboration~\cite{CrystalBarrel:1997kda} performed
proton annihilation to produce the $K\bar{K}\pi$ system and discovered
the $h_1$ meson above the $K\bar{K}^*$ threshold, with a mass of
$M=(1440\pm 60)$ MeV and a width of $\Gamma=(170\pm 80)$ MeV. Two
experimental results from the BESIII Collaboration, obtained from
different charmonium state decays, confirmed these findings with a
slightly lower mass. It is noteworthy that interference influences the
determination of the mass and width of the
$h_1(1415)$~\cite{BESIII:2018ede}, and these results align 
with those of the Crystal Barrel experiment. 
In a recent experiment conducted by the
BESIII~\cite{BESIII:2022zel} Collaboration, the mass and width of the
$h_1(1415)$, measured in the $\gamma\eta'$ invariant mass, were found
to be consistent with those obtained by the LASS Collaboration but
contradictory to the two previous BESIII results. This discrepancy
strongly suggests that the $h_1(1415)$ does not originate from a
single pole, indicating a more complex structure. Evidently, this
structure cannot be explained by the quark model alone. Consequently,
based on the results of this study, we propose a two-pole structure to
account for the discrepancy in the measurements of the $h_1(1415)$
meson mass. 

\begin{figure}[htbp]
	\centering
	\includegraphics[scale=0.9]{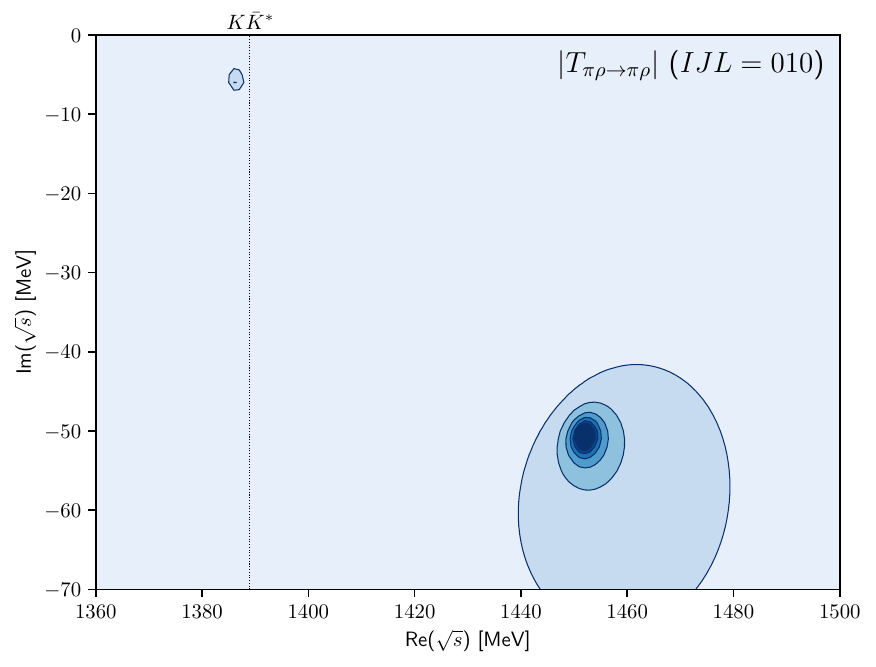}
	\caption{Contour plot of $\pi\rho\to\pi\rho$ transition
          amplitude  in the complex $\sqrt{s}$ plane, which enlarges
          the red rectangle in Fig.~\ref{fig:5}.}
	\label{fig:7}
\end{figure}
The current work reveals two poles around 1.4 GeV, one at $(1387
-i6)$ MeV and the other at $(1452-i51)$ MeV, as shown in
Table~\ref{tab:2}. Figure~\ref{fig:7} demonstrates how these two poles
appear in the complex energy plane. The lower pole is located close to
the $K\bar{K}^*$ threshold, while the higher pole is situated 80 MeV
above it. Due to its narrowness, detecting the lower pole
experimentally may pose a challenge. 
Examining the coupling strengths in Table~\ref{tab:2}, we observe that
both poles couple strongly to open strange and hidden strange
channels. However, the higher pole couples to the $\eta\phi$ channel
far more strongly than to other channels. This suggests that the
higher pole might be an $\eta\phi$ molecular state, while the lower
pole, being located very close to the $K\bar{K}^*$ threshold, could be
a $K\bar{K}^*$ molecular state. 
Notably, the $g_{K\bar{K}^*}$ for the higher pole has a larger
imaginary part than its real part, causing destructive
interference. This results in the disappearance of the higher pole in
$K\bar{K}^*$ elastic scattering, a feature also observed in the case of
the higher pole of the $b_1$ meson in a previous
study~\cite{Clymton:2023txd}. This characteristic is the essential
clue to explaining the absence of the higher pole in 
the LASS Collaboration experiment. 

Based on this analysis, we propose an explanation for the conflicting 
mass measurements of $h_1(1415)$. As shown in Fig.~\ref{fig:6}, the
crucial aspect is that the higher pole vanishes in $K\bar{K}^*$ elastic
scattering, leaving only the $K\bar{K}^*$ threshold enhancement. The
LASS collaboration investigated the $h_1(1415)$ in the $K^-p\to
K_S^0\bar{K}\pi\Lambda$ process, which can effectively be represented
as $K\bar{K}^*$ elastic scattering. They observed a threshold
enhancement near the $K\bar{K}^*$ threshold, aligning with the current
results. In contrast, the Crystal Barrel~\cite{CrystalBarrel:1997kda}
and BESIII experiments~\cite{BESIII:2015vfb, BESIII:2018ede} measured
the $K\bar{K}\pi$ state as a final state, which can be considered as
the $K\bar{K}^*$ state. Consequently, both collaborations measured the 
$h_1(1415)$ above the $K\bar{K}^*$ threshold. The most recent
measurement from the BESIII Collaboration comes from the $J/\psi\to
\gamma \eta'\eta'$ decay. 
In conclusion, we propose that the $h_1(1415)$ can be constructed from
two different resonances: the $h_1(1380)$ and $h_1(1440)$ mesons, with
the higher pole vanishing in $K\bar{K}^*$ elastic scattering. This
two-pole structure accounts for the seemingly conflicting experimental
results. 

\section{Summary and conclusions}
In this work, we aimed at investigating the isoscalar axial-vector
$h_1$ mesons using a coupled-channel formalism. We first constructed
kernel amplitudes from meson-exchange diagrams in the $t$- and
$u$-channels, derived from effective Lagrangians based on hidden local
symmetry. In addition, we introduced the pole diagram for
$h_1(1595)$ to generate the $h_1(1595)$ resonance. We found that 
the $h_1(1595)$ pole diagram also has a significant effect on the
generation of $h_1(1170)$. Six channels were incorporated: $\pi\rho$,
$\eta\omega$, $K\bar{K}^*$, $\eta\phi$, $\eta'\omega$, and $\eta'\phi$
channels. We solved the off-shell coupled integral equations and
discussed the dynamical generation of the $h_1(1170)$.  
The present analysis revealed two poles at $(1387-i6)$
MeV and $(1452-i51)$ MeV, exhibiting a two-pole structure of the
$h_1(1415)$ meson. This two-pole structure may resolve the discrepancy
in the experimental data on the mass of $h_1(1415)$. The results
showed that the lower pole couples strongly to the $K\bar{K}^*$
channel, while the higher pole couples predominantly to the $\eta\phi$
channel. This provides an essential clue to understand the nature of
$h_1$ mesons and explains possible discrepancies in the mass of
$h_1(1415)$. The two-pole structure of $h_1(1415)$ can account for the
seemingly conflicting experimental results, with the lower pole
corresponding to $h_1(1380)$ and the higher pole to
$h_1(1440)$. Notably, the higher pole vanishes in $K\bar{K}^*$ elastic
scattering, which explains why some experiments observe only the lower
pole.  
 
\begin{acknowledgments}
  The authors are grateful to Terry Mart for valuable discussions. Part of
  the work was done at Department of Physics, University of Indonesia.
  HCK wants to express his gratitude to Tetsuo Hyodo, Makoto Oka, and
  Qian Wang for valuable discussions. The present work is supported by
  Inha University Research Grant in 2024 (No.73014-1).   
\end{acknowledgments}

\bibliography{h1meson}
\bibliographystyle{apsrev4-2}

\end{document}